\documentclass[12pt]{article}

\usepackage{amsmath}
\usepackage{amssymb}
\usepackage{amsfonts}
\usepackage{graphicx}

\numberwithin{equation}{section} 

\begin{document}

\begin{center}
{\large \bf{  Confronting  the Ellipsoidal Universe to the Planck 2018 Data~\footnote{Based on observations obtained 
with Planck (http://www.esa.int/Planck), an ESA science mission with instruments and contributions directly funded by ESA Member States, 
NASA, and Canada.}  } }
\end{center}

\vspace*{2cm}

\begin{center}
{
Paolo Cea~\protect\footnote{Electronic address:
{\tt paolo.cea@ba.infn.it}}  \\[0.5cm]
{\em INFN - Sezione di Bari, Via Amendola 173 - 70126 Bari,
Italy} }
\end{center}

\vspace*{1.5cm}

\begin{abstract}
\noindent 
We compared the Planck 2018 data on the large-scale anisotropies in the cosmic microwave background 
to  the results obtained in  the slightly anisotropic  ellipsoidal universe.  
We focused on the quadrupole temperature correlations to better constrain the 
 eccentricity at decoupling and the direction of the symmetry axis. We found that  the quadrupole TE and EE  temperature correlations
 in  the  ellipsoidal universe are still in reasonable  agreement with the Planck 2018 data. We suggested that an experimental  estimate
 of the average large-scale polarisation  by the Planck Collaboration could   confirm or reject the anisotropic  universe
 proposal.
\vspace{0.8cm}

\noindent
{\bf keywords}: cosmic microwave radiation;  cosmology: theory

\vspace{0.2cm}
\end{abstract}

\newpage

\noindent
\section{Introduction}
\label{S1}
The temperature  anisotropies in the cosmic microwave background (CMB) is one of the most powerful way to study cosmology and
the physics of the early Universe. Recently, the Planck Collaboration reported the final results on the CMB anisotropies 
\cite{Aghanim:2018a,Aghanim:2018b,Aghanim:2018c,Aghanim:2018d,Akrami:2018a,Akrami:2018b,Akrami:2018c,Akrami:2018d,Akrami:2018e}
confirming the cosmological  Lambda Cold Dark Matter ($\Lambda$CDM) model to the highest level of accuracy. 
Nevertheless, even the Planck 2018 data are confirming the presence of anomalous features at  large scales.
The most evident anomaly concerns the quadrupole temperature correlation that is still heavily suppressed with respect to the best-fit
standard cosmological  $\Lambda$CDM model.  There are  several proposals in the literature to cope with the suppression of power at large scales in the CMB anisotropies. In particular, it has been suggested~\cite{Campanelli:2006,Campanelli:2007} that, 
if one admits that the large-scale spatial geometry of our universe could be only  plane-symmetric,
then the quadrupole amplitude can be drastically reduced without affecting higher multipoles of the angular power spectrum of the temperature anisotropies. If this is the case, then  the metric of the  Friedmann-Robertson-Walker (FRW) standard cosmological model 
(see, for instance, Ref.~\cite{Peebles:1993})~\footnote{We shall use throughout the natural units  $c \; = \; 1$, $\hbar \; = \; 1$,  $k_B \; = \; 1$.}:
\begin{equation}
\label{1.1}
ds^2 = - dt^2 + a^2(t) \delta_{ij}  \, dx^i dx^j \; ,
\end{equation}
should be  replaced with the ellipsoidal universe metric: 
\begin{equation}
\label{1.2}
ds^2 = - dt^2 + a^2(t) (\delta_{ij} + h_{ij}) \, dx^i dx^j \; ,
\end{equation}
where $h_{ij}$ is a metric perturbation:
\begin{equation}
\label{1.3}
h_{ij} \; = \;  - \, e^2(t) \;  n_{i}  \, n_{j} \; ,
\end{equation}
where $e(t)$ is the ellipticity and the unit vector $\vec{n}$ determines the direction of the symmetry axis.  In fact, the ellipsoidal universe
corresponds to a Bianchi type I~\cite{Bianchi:1898} cosmological model with planar symmetry. It is known since long 
time~\cite{Rees:1968, Negroponte:1980,Basko:1980,Pontzen:2007} that anisotropic cosmological models in general
induce at large scales  sizeable E- and B-mode polarisation in  the the cosmic microwave background radiation. 
Actually,  the recent results~\cite{Ade:2015} of a joint analysis of CMB data from BICEP2/Keck Array and Planck Collaborations found
no statistically significant evidence of primordial B-modes. This should select Bianchi type I cosmological models as viable anisotropic 
cosmological models since B-mode polarisation is being produced in all Bianchi types except type I. 
Indeed, in our previous papers~\cite{Cea:2010,Cea:2014} we showed that the ellipsoidal geometry of the universe could induce
sizeable linear polarisation signal at large scales without invoking reionization processes. In particular, in Ref.~\cite{Cea:2014}
we evaluated the quadrupole TE and EE correlations and compared with the WMAP nine-year data~\cite{Bennett:2013,Hinshaw:2013}.
The aim of the present paper is to critically  contrast the CMB quadrupole correlations evaluated  in the ellipsoidal universe with
the recent Planck 2018 data. \\
The plan of the paper is as follows. In sect.~\ref{S2}  we  summarise the main results presented in Ref.~\cite{Cea:2014};
sect.~\ref{S3} is devoted to the calculations of the quadrupole correlations and, by using the Planck 2018 data,
 to better constrain the eccentricity at decoupling and the symmetry axis. We also compare in sect.~\ref{S3.1} the quadrupole
 TE and EE correlations to the Planck data. Finally, in sect.~\ref{S4}  we summarise the main results of the present paper
 and we draw  our conclusions.
\section[]{The CMB large scale anisotropies}
\label{S2}
In this section, for later  convenience, we summarise the calculations of the large scale temperature anisotropies in the 
 ellipsoidal universe~\cite{Campanelli:2006,Campanelli:2007}  presented in Ref.~\cite{Cea:2014}. \\
Let us consider the Boltzmann equation for the photon distribution in the ellipsoidal universe
 by taking into account also the effects of the cosmological  inflation produced primordial scalar perturbations.
The temperature anisotropies caused by the  inflation produced primordial scalar perturbations
are discussed in  textbooks (see, for instance,  Refs.~\cite{Dodelson:2003}, \cite{Mukhanov:2005}).
Here, we consider the  primordial scalar perturbations induced by the inflation. In the conformal Newtonian gauge~\cite{Mukhanov:2005}
the effects of these perturbations on the metrics are accounted for by two functions  $\Psi (\vec{x},t)$ and  $\Phi (\vec{x},t)$
corresponding to the Newtonian potential and the perturbation to the spatial curvature, respectively. Therefore, in the
ellipsoidal universe the perturbed metrics reads:
\begin{equation}
\label{2.1}
ds^2  =  - [1 \, + \, 2 \Psi (\vec{x},t)] \, dt^2  \;  + \; a^2(t)  \; \{  \delta_{ij} \,  [1 \, + \, 2 \Phi (\vec{x},t)] \, + \,  h_{ij} \} \,   dx^i dx^j \; .
\end{equation}
We need to evaluate the temperature fluctuations of the cosmic background radiation induced by eccentricity of the universe and by 
the inflation produced primordial cosmological perturbations.  We assume that  the photon distribution function $f(\vec{x},t)$ is an 
isotropically radiating blackbody at a sufficiently early epoch.  The subsequent evolution of  $f(\vec{x},t)$ is determined by the Boltzmann  
 equation~\cite{Dodelson:2003,Mukhanov:2005}:
\begin{equation}
\label{2.2}
\frac{df}{dt} \; = \; \left ( \frac{\partial f} {\partial t} \right )_{coll} \; , 
\end{equation}
where $ (\frac{\partial f} {\partial t} )_{coll} $ is the collision integral which takes care of Thomson scatterings between matter and radiation.
The distribution function depends on the space-time point $x^{\mu}$ and the momentum vector   $p^{\mu}$: 
\begin{equation}
\label{2.3}
 p^{\mu} \;  =  \;  \frac{d \, x^{\mu}}{d \, \lambda} \; ,
\end{equation}
where $\lambda$ parametrises the particle's path.  Actually, we may consider the distribution function as a function of 
the magnitude of momentum $p$ and  the momentum direction $\hat{p}^i$, $ \delta_{ij} \,  \hat{p}^i \hat{p}^j = 1$.
Tacking into account the metric Eq.~(\ref{2.1})  one gets~\cite{Cea:2014}:
\begin{equation}
\label{2.4}
\frac{df}{dt} \; \simeq \; \frac{\partial f}{\partial t}  \, +  \, \frac{\hat{p}^i}{a(t)} \, \frac{\partial f}{\partial x^i} \, - p \,  \frac{\partial f}{\partial p} \,  
 [ H(t) \, + \,    \frac{\partial \Phi}{\partial t}  \, + \,  \frac{\hat{p}^i}{a(t)}  \, \frac{\partial \Psi}{\partial x^i}  \, + \, 
  \frac{1}{2}   \hat{p}^{i} \, \hat{p}^{j}    \frac{\partial h_{i  j} }{\partial t}  ] \; ,
\end{equation}
where  $H = \dot{a}/a$ is the Hubble rate. Now,  the  photon distribution perturbed with respect to
  its zero-order Bose-Einstein value:
\begin{equation}
\label{2.5}
f_0(p,t) \; =  \;  \frac{1}{e^{\frac{p}{T(t)}} \, - \, 1} \; ,
\end{equation}
can be written as:
\begin{equation}
\label{2.6}
f(\vec{x},t,p,\hat{p}) =  \frac{1}{e^{\frac{p}{T(t)[1\, + \, \Theta(\vec{x},t,p,\hat{p})]}} \, - \, 1} \; .
\end{equation}
After expanding to the first order in the perturbation  $\Theta(\vec{x},t,p,\hat{p})$ we get:
\begin{equation}
\label{2.7}
f(\vec{x},t,p,\hat{p}) \; \simeq \; f_0(p,t)  \; [ 1 \; + \;  \Theta(\vec{x},t,p,\hat{p}) ]  \; .
\end{equation}
Note that from Eq.~(\ref{2.6}) it follows that the distribution function   $\Theta(\vec{x},t,p,\hat{p})$ is  the temperature contrast function:
\begin{equation}
\label{2.8}
\Theta(\vec{x},t,p,\hat{p}) \; = \; \frac{\Delta T(\vec{x},t,p,\hat{p})}{T(t)} \; . 
\end{equation}
The perturbed distribution $\Theta(\vec{x},t,p,\hat{p})$ can be obtained by solving the Boltzmann equation to the first order.  
In this approximation we obtain:
\begin{equation}
\label{2.9}
\frac{\partial \Theta }{\partial t}  \, +  \, \frac{\hat{p}^i}{a(t)} \, \frac{\partial \Theta}{\partial x^i} \, 
 + \,    \frac{\partial \Phi}{\partial t}  \, + \,  \frac{\hat{p}^i}{a(t)}  \, \frac{\partial \Psi}{\partial x^i}  \, + \, 
  \frac{1}{2}   \hat{p}^{i} \, \hat{p}^{j}    \frac{\partial h_{i  j} }{\partial t}   \; \simeq  \;  
 \frac{1}{f_0}  \;     \left ( \frac{\partial f} {\partial t} \right )_{coll}  \; . 
\end{equation}
In the same approximations the collision integral is a linear functional of $\Theta(\vec{x},t,p,\hat{p})$.
 Moreover, we showed~\cite{Cea:2014} that  at large scales  the collision integral can be considered a linear
homogeneous functional of the distribution function  $\Theta(\vec{x},t,p,\hat{p})$.  Therefore, writing:
\begin{equation}
\label{2.10}
\Theta(\vec{x},t,p,\hat{p})    \;  \simeq \; \Theta^A(\vec{x},t,p,\hat{p}) \; + \;  \Theta^I(\vec{x},t,p,\hat{p})  \; ,
\end{equation}
we have:
\begin{equation}
\label{2.11}
 \left ( \frac{\partial f} {\partial t} \right )_{coll}[\Theta] \; \simeq \;
 \left ( \frac{\partial f} {\partial t} \right )_{coll}[\Theta^A] \; + \;   \left ( \frac{\partial f} {\partial t} \right )_{coll}[\Theta^I]  \; .
\end{equation}
In fact, Eqs.~(\ref{2.10}) and (\ref{2.11})   show  that   $\Theta^A$ and $\Theta^I$ are  the temperature fluctuations 
induced by the spatial anisotropy of the geometry of the universe and by the scalar perturbations
generated during the inflation, respectively.  Accordingly we obtain:
\begin{equation}
\label{2.12}
\frac{\partial \Theta^I }{\partial t}  \, +  \, \frac{\hat{p}^i}{a(t)} \, \frac{\partial \Theta^I}{\partial x^i} \, 
 + \,    \frac{\partial \Phi}{\partial t}  \, + \,  \frac{\hat{p}^i}{a(t)}  \, \frac{\partial \Psi}{\partial x^i}    \; \simeq  \;  
 \frac{1}{f_0}  \;     \left ( \frac{\partial f} {\partial t} \right )_{coll}[\Theta^I]  \; , 
\end{equation}
and
\begin{equation}
\label{2.13}
\frac{\partial \Theta^A }{\partial t}  \, +  \, \frac{\hat{p}^i}{a(t)} \, \frac{\partial \Theta^A}{\partial x^i} \, 
 + \,   \frac{1}{2}   \hat{p}^{i} \, \hat{p}^{j}    \frac{\partial h_{i  j} }{\partial t}      \; \simeq  \;  
 \frac{1}{f_0}  \;     \left ( \frac{\partial f} {\partial t} \right )_{coll}[\Theta^A]  \; .
\end{equation}
It is worthwhile to stress that our results imply that at large scales:
\begin{equation}
\label{2.14}
\Delta T(\vec{x},t,p,\hat{p})  \; \simeq \;   \Delta T^A(\vec{x},t,p,\hat{p}) \; + \;  \Delta T^I(\vec{x},t,p,\hat{p}) \; ,
\end{equation}
where   $\Delta T^I$ and   $ \Delta T^A$ are the temperature fluctuations  induced by the cosmological scalar perturbations and by the spatial anisotropy of the metric of the universe. \\
 To determine the CMB temperature fluctuations at large scales we need to solve the Boltzmann
equations  Eqs.~(\ref{2.12}) and  (\ref{2.13}).  Eq.~(\ref{2.12}) is the Boltzmann equation of the standard $\Lambda$CDM cosmological
model, and it has been extensively discussed in literature.  As a consequence, we only need to solve
 the Boltzmann equation  Eq.~(\ref{2.13}) which allows us to find the CMB temperature fluctuations caused by the anisotropy of the geometry
of the universe. To this end,  we introduce the Fourier transform of the  temperature contrast function:
\begin{equation}
\label{2.15}
 \Theta^A(\vec{x},t,p,\hat{p})  \; = \;   \int \, \frac{d^3k}{(2 \pi)^3} \; e^{i \, \vec{k} \, \cdot \,  \vec{x} }  \;  \Theta^A(\vec{k},t,p,\hat{p})  \; .
\end{equation}
Considering that the collision integral depends linearly on  $\Theta^A$, we have:
\begin{equation}
\label{2.16}
\frac{\partial \Theta^A(\vec{k},t,p,\hat{p})   }{\partial t}  \, +  \, \frac{i \, \vec{k} \cdot \hat{p}}{a(t)} \, \Theta^A(\vec{k},t,p,\hat{p}) \, 
 + \,   \frac{1}{2}   \hat{p}^{i} \, \hat{p}^{j}    \frac{\partial h_{i  j} }{\partial t}      \; \simeq  \;  
 \frac{1}{f_0}  \;     \left ( \frac{\partial f} {\partial t} \right )_{coll} [\Theta^A(\vec{k},t,p,\hat{p})]  \; .
\end{equation}
To determine the polarisation of the cosmic microwave background we need the polarised distribution function which, in general, is represented by a column vector whose components are the four Stokes parameters~\cite{Chandrasekhar:1960}. In fact, due to the axial symmetry of the metric only two Stokes parameters need to be considered, namely the two intensities of radiation with electric vectors  in the plane containing $\vec{p}$ and $ \vec{n}$  and perpendicular to this plane respectively. As a consequence,   Eq.~(\ref{2.7})  is replaced by:
\begin{equation}
\label{2.17}
f(\vec{x},t,p,\hat{p}) \; \simeq \; f_0(p,t)  \;  \left[ \begin{pmatrix} 1 \\ 1 \end{pmatrix}  \; + \; \Theta^A(\vec{x},t,p,\hat{p})  \right] \; \; ,
\end{equation}
where, now,  $ \Theta^A(\vec{x},t,p,\hat{p})$  should be regarded as a  two component column vector.  
Defining 
\begin{equation}
\label{2.18}
\mu \;  = \;  \cos \theta_{\vec{p}  \vec{n}} \; \; , \; \;  \cos \theta_{\vec{k} \vec{ p}} \; = \;    \frac{\vec{k} \cdot \hat{p}}{k} \; ,
\end{equation}
we get from  Eq.~(\ref{2.16}):
\begin{eqnarray}
\label{2.19}
&& \frac{\partial \Theta^A(\vec{k},t,\mu)   }{\partial t}  \, + \,  \frac{i \, k}{a(t)}   \cos \theta_{\vec{k} \vec{ p}}   \; \Theta^A(\vec{k},t,\mu) \;
 \simeq  \; \frac{1}{2} \;  \left [ \frac{d}{d \, t} \, e^2(t) \right ] \;  \mu^2  \;  \begin{pmatrix} 1 \\  1 \end{pmatrix} \nonumber \\
&& - \, \sigma_T \, n_e  \left [ \Theta^A(\vec{k},t, \mu) \,   - \,  \frac{3}{8} \int_{-1}^1
\begin{pmatrix} 2(1-\mu^2)(1-\mu^{\prime 2})+\mu^2 \mu^{\prime 2} & \mu^2 \\
\mu^{\prime 2} & 1 \end{pmatrix}  \,  \Theta^A(\vec{k},t, \mu') \; d\mu' \right] 
\nonumber \\
\end{eqnarray}
where $\sigma_T$ is the Thomson cross section and $ n_e(t)$  the electron number density ~\cite{Chandrasekhar:1960}. 
After introducing the conformal time:
\begin{equation}
\label{2.20}
\eta(t) \; =  \;  \int_{0}^t \frac{dt'}{a(t')} \; \; ,
\end{equation}
 Eq.~(\ref{2.19}) can be written as:
\begin{eqnarray}
\label{2.21}
&& \frac{\partial \Theta^A(\vec{k},\eta,\mu)   }{\partial \eta}  \, + \,  i \, k \;  \cos \theta_{\vec{k} \vec{ p}}   \; \Theta^A(\vec{k},\eta,\mu) \;
 \simeq  \; \frac{1}{2} \;  \left [ \frac{d}{d \, \eta} \, e^2(\eta) \right ] \; ( \mu^2 \, - \, \frac{1}{3})  \;  \begin{pmatrix} 1 \\ 1 \end{pmatrix} \nonumber \\
&& -  a(\eta) \, \sigma_T \, n_e \left [ \Theta^A(\vec{k},\eta, \mu) \,   - \,  \frac{3}{8} \int_{-1}^1
\begin{pmatrix}  2(1-\mu^2)(1-\mu^{\prime 2})+\mu^2 \mu^{\prime 2} & \mu^2 \\
\mu^{\prime 2} &1 \end{pmatrix}  \,  \Theta^A(\vec{k},\eta, \mu') \; d\mu' \right]  \; . 
\nonumber \\
\end{eqnarray}
The  solutions of Eq.~(\ref{2.21}) are given by~\cite{Basko:1980,Cea:2010,Cea:2014}:
\begin{equation}
\label{2.22}
\Theta^A(\vec{k},\eta,\mu)  \; =  \;   \theta_a(\vec{k},\eta) \, (\mu^2 - \frac{1}{3})   \begin{pmatrix} 1 \\ 1 \end{pmatrix} \; + 
 \;   \theta_p(\vec{k},\eta) \, (1 - \mu^2 )    \begin{pmatrix} 1 \\ -1 \end{pmatrix} \; .
\end{equation}
It turns out that   $  \theta_a(\vec{k},\eta)$ measures the degree of anisotropy, while $\theta_p(\vec{k},\eta)$ gives the polarisation 
of the primordial radiation.
In Ref.~\cite{Cea:2014} we found:
\begin{equation}
\label{2.23}
  \theta_a(\vec{k},\eta) = \frac{1}{7}  \int_{\eta_i}^{\eta} \Delta H(\eta') \left[ 6 e^{-\tau(\eta,\eta')} \,  +  \, e^{- \frac{3}{10} \tau(\eta,\eta')}  \right] 
  \;  e^{i \, k \,   \cos \theta_{\vec{k} \vec{ p}} (\eta' - \eta)}  \;  d\eta'   \; ,
\end{equation}
\begin{equation}
\label{2.24}
  \theta_p(\vec{k},\eta) = \frac{1}{7}  \int_{\eta_i}^{\eta} \Delta H(\eta') \left[  e^{-\tau(\eta,\eta')} \,  - \, e^{- \frac{3}{10} \tau(\eta,\eta')}  \right] 
  \;  e^{i \, k \,   \cos \theta_{\vec{k} \vec{ p}} (\eta' - \eta)}  \;  d\eta'   \; ,
\end{equation}
where  we introduced the  cosmic shear~\cite{Negroponte:1980,Cea:2010,Cea:2014}:
\begin{equation}
\label{2.25}
\Delta H(\eta) \;  \equiv  \;  \frac{1}{2} \; \frac{d }{d \eta}e^2(\eta)  \; \; ,
\end{equation}
and the optical depth:
\begin{equation}
\label{2.26}
 \tau(\eta,\eta') =   \int_{\eta'}^{\eta} \sigma_T \, n_e \, a(\eta'') \, d\eta'' \; \; .
\end{equation}
To a good approximation,   we showed that~\cite{Cea:2014}:
\begin{equation}
\label{2.27}
\Theta(\vec{k},\eta_0,\mu, \hat{p})  \; \simeq \;   \theta_a \, (\mu^2 - \frac{1}{3})  \;  e^{- \, i \, k \,   \cos \theta_{\vec{k} \vec{ p}} \,  \eta_0} \;  \begin{pmatrix} 1 \\ 1  \end{pmatrix} \; +  \;  
 \theta_p \, (1 - \mu^2 ) \;  e^{- \, i \, k \,   \cos \theta_{\vec{k} \vec{ p}} \,  \eta_0} \;    \begin{pmatrix} 1 \\ -1 \end{pmatrix} \; ,
\end{equation}
where $\eta_0$ is the conformal time at the present time. Moreover, we also found:

\begin{equation}
\label{2.28}
\theta_p \;  \simeq \;  8.92 \; 10^{-3} \;  e^2_{\rm dec}  \;   \; .
\end{equation}
and
\begin{equation}
\label{2.29}
\theta_a \;  =  \;  - \; \frac{1}{2} \;  \epsilon \; \;  \; , \; \; \;  \epsilon \; \simeq \;  0.944  \;  e^2_{\rm dec}  \; .
\end{equation}
where  $e_{\rm dec}$ is the ellipticity at decoupling.
\section[]{The quadrupole correlations}
\label{S3}
The temperature anisotropies of the cosmic background depend on the polar angle $\theta, \phi$, so that one usually expands in terms of 
 spherical harmonics:
\begin{equation}
\label{3.1}
\Delta T(\theta,\phi) \;  = \; 
\sum_{\ell = 1 }^{\infty} \sum_{m= - \ell}^{+ \ell} \; a_{\ell m} \, Y_{\ell m}(\theta,\phi)  \; .
\end{equation}
 The CMB temperature fluctuations can be fully characterised by the power spectrum:
\begin{equation}
\label{3.2}
( \Delta T_{\ell} )^2  \;  \equiv  \; \mathcal{D}_{\ell}   \; = \; \frac{\ell(\ell+1)}{2 \pi}  \; C_{\ell} \; \; , \; \; 
C_{\ell} \; = \; \frac{1}{2 \ell+1} \sum_{m=-\ell}^{+\ell} \,  | a_{\ell m} |^2 \; .
\end{equation}
In particular, the quadrupole anisotropy refers to the multipole $\ell = 2$. Remarkably, also the Planck 2018 data~\cite{Planck:2018} 
confirmed that the observed quadrupole anisotropy:
\begin{equation}
\label{3.3}
( \Delta T_{2} )^2  \;  =  \; \mathcal{D}_{2}   \; \simeq \; 
225.9 \;  \;  \mu \, K^2 \;  , 
\end{equation}
is much smaller than the quadrupole anisotropy expected  according to the ' TT,TE, EE + low E + lensing ' best fit  $\Lambda$CDM model to the Planck 2018 data~\cite{Planck:2018}:
\begin{equation}
\label{3.4}
( \Delta T^I_{2} )^2  \;  =  \;  1017 \; \pm \; 643 \;   \mu \, K^2 \; .
\end{equation}
Note that in Eq.~(\ref{3.4})  the uncertainty is  due to the so-called cosmic variance~\cite{Dodelson:2003} 
that, in fact,  should be included in the theoretical expectations.  \\
In the standard cosmological model the CMB temperature fluctuations are induced by the cosmological perturbations of
the  FRW homogeneous and isotropic background metric generated by the inflation-produced potentials.
In the ellipsoidal universe we must also consider the effects on the CMB anisotropies induced
by the anisotropic expansion of the universe. It turned out~\cite{Cea:2010,Cea:2014} that, as discussed in sec.~\ref{S2}, 
at large scales the observed anisotropies in the CMB temperature are due to the linear superposition of the two contributions according
to  Eq.~(\ref{2.14}).
Following Ref.~\cite{Cea:2014}, let us introduce the dimensionless  temperature anisotropies: 
\begin{equation}
\label{3.5}
\frac{\Delta T(\theta,\phi)}{ T_0} \;  = \; 
\sum_{\ell = 1 }^{\infty} \sum_{m= - \ell}^{+ \ell} \; a_{\ell m} \, Y_{\ell m}(\theta,\phi)  \; ,
\end{equation}
where  $T_0 \simeq 2.7255 \; K$~\cite{Fixsen:2009} is the actual (average) temperature of the CMB radiation.
Obviously,  the $a_{\ell m}$'s  in  Eq.~(\ref{3.5}) are dimensionless and can be  obtained from
the corresponding coefficients in  Eq.~(\ref{3.1})  by dividing by $T_0$.   After that, one introduces the power spectrum:
\begin{equation}
\label{3.6}
( \frac{\Delta T_{\ell}}{ T_0 } )^2 \; =  \;  \frac{1}{2 \pi} \,
\frac{\ell (\ell+1)}{2 \ell + 1} \sum_m | a_{\ell m} |^2 \; ,
\end{equation}
that fully characterises the properties of the CMB temperature anisotropy.  We have seen that at large scales the temperature
fluctuations $ \Delta T$ are the sum of  the temperature fluctuations $\Delta T^I$ induced by the cosmological inflation perturbations  and 
 the temperature fluctuations   $ \Delta T^A$ due to the spatial anisotropy of the metric of the universe.  Writing:
\begin{equation}
\label{3.7}
\frac{\Delta T^I(\theta,\phi)}{ T_0} \;  = \; 
\sum_{\ell = 1 }^{\infty} \sum_{m= - \ell}^{+ \ell} \; a^I_{\ell m} \, Y_{\ell m}(\theta,\phi)  \; ,
\end{equation}
and
\begin{equation}
\label{3.8}
\frac{\Delta T^A(\theta,\phi)}{ T_0} \;  = \; 
\sum_{\ell = 1 }^{\infty} \sum_{m= - \ell}^{+ \ell} \; a^A_{\ell m} \, Y_{\ell m}(\theta,\phi)  \; ,
\end{equation}
we have:
\begin{equation}
\label{3.9}
a_{\ell m} \; =  \; a_{\ell m}^{ A} \;  + \; a^{ I}_{\ell m} \; .
\end{equation}
In our previous paper~\cite{Cea:2014} we argued that the main contributions of the anisotropy of the metric  to the  CMB temperature anisotropies are for $\ell=2$.  Therefore  we are led to consider the quadrupole anisotropy $\ell=2$.
Firstly, let us consider  the contributions to  temperature contrast function induced by the   anisotropic expansion of the universe.
Since at large scales we may neglect the spatial dependence of the  contrast function, we may set $k \, \simeq \, 0$
in   Eq.~(\ref{2.27}).  In this case we obtain at once:
\begin{equation}
\label{3.10}
\frac{\Delta T^{A}(\theta,\phi)}{ T_0} \; \simeq \; \Theta^{A}(\vec{k} \simeq \, 0 ,\eta_0,\mu, \hat{p}) 
= \;  \theta_a \, ( \cos^2 \theta_{\vec{p}  \vec{n}}  - \frac{1}{3})  \;  ,    
\end{equation}
where $\theta_a$ is given by  Eq.~(\ref{2.28}) and $\theta,\phi$ are the polar angles of the photon momentum $\vec{p}$. 
Let $\theta_n, \phi_n$ be the polar angles of the direction of the axis of symmetry $\vec{n}$, then:
\begin{equation}
\label{3.11}
\frac{\Delta T^{A}(\theta,\phi)}{ T_0} \;  \simeq \; \frac{2}{3} \;  \theta_a \; P_2( \cos \theta_{\vec{p}  \vec{n}})   \;  
= \; \frac{2}{3} \;  \theta_a \;  \frac{ 4 \pi}{5} \, \sum_{m= - 2}^{+ 2} \;  \, Y_{2 m}(\theta,\phi)  \, Y^*_{2 m}(\theta_n,\phi_n)  \;  .    
\end{equation}
As a consequence, we get: 
\begin{equation}
\label{3.12}
a_{2 m}^{ A}  \; \simeq \;   - \,  \frac{ 4 \pi}{15}  \, \epsilon^2   \;  Y^*_{\ell m}(\theta_n,\phi_n)  \;  \; , \; \; 
 a_{\ell m}^{ A} = 0  \; \;  {\text for} \; \;   \ell \neq 2 \; .
\end{equation}
From Eq.~(\ref{3.12}) one easily finds:
\begin{eqnarray}
\label{3.13}
&& a_{20}^{A} \; \simeq \;  + \, \frac{ 1}{6}  \, \epsilon^2   \sqrt{ \frac{\pi}{5} } \,
                  [1 + 3 \cos^2 (2 \theta_n) ] \;  ,  \nonumber \\
&& a_{21}^{ A} \;  = \;  (a_{2,-1}^{A})^{*} \; \simeq \; + \, i \, 
                  \sqrt{  \frac{\pi}{30} }  \, \epsilon^2 
                            e^{-i \phi_n}  \sin (2 \theta_n )  \;  , \nonumber \\
&& a_{22}^{ A} \; = \; (a_{2,-2}^{ A})^{*} \;  \simeq \; + \,
                   \sqrt{  \frac{ \pi}{30} } \, \epsilon^2   
                  \; e^{-2 i \phi_n}  \sin^2 \theta_n  \; .
\end{eqnarray}
After defining the quadrupole anisotropy:
\begin{equation}
\label{3.14}
\mathcal{Q}_A^2  \;  \equiv  \, ( \frac{\Delta T^A_2}{ T_0} )^2  \;  ,
\end{equation}
one finds:
\begin{equation}
\label{3.15}
\mathcal{Q}_{ A} \;  \simeq \; \frac{2}{5 \sqrt{3} } \; \epsilon^2  \;  .
\end{equation}
Concerning  the observed quadrupole temperature anisotropy, we introduce:
\begin{equation}
\label{3.16}
\mathcal{Q}^2  \;  \equiv  \, ( \frac{\Delta T_2}{ T_0} )^2  \;  ,
\end{equation}
so that, according to Eq.~(\ref{3.9}), we have:
\begin{equation}
\label{3.17}
\mathcal{Q}^2   \; =  \;  \frac{3}{5 \,  \pi} \;
 \sum_{m=-2}^{m=+2}  | a^A_{2m} \; + \;  a^I_{2m} |^2 \;  \; . 
\end{equation}
We see that to determine $\mathcal{Q}^2$   we need the $a^I_{\ell m}$' s.  Since the standard inflation-produced temperature
fluctuations are statistically isotropic,  we can write~\cite{Campanelli:2007,Cea:2014}:
\begin{eqnarray}
\label{3.18}
&& a^{I}_{20} \; \simeq \;  \sqrt{\frac{\pi}{3}} \;   \mathcal{Q}_{I}, \nonumber  \\
&& a^{I}_{21} \; = \;  - \,  (a^{\rm I}_{2,-1})^{*} \; \simeq \; + \, i \,  \sqrt{\frac{\pi}{3}} \; e^{i \phi_1} \;  \mathcal{Q}_{ I}  \; , \\
&& a^{ I}_{22} \; =  \; (a^{\rm I}_{2,-2})^{*} \;  \simeq \; \sqrt{\frac{\pi}{3}}
\; e^{i \phi_2}  \; \mathcal{Q}_{I}  \;  , \nonumber
\end{eqnarray}
where $0 \leq \phi_1, \phi_2  \leq 2 \pi$ are unknown phases and  $ \mathcal{Q}_{I}$, defined by
\begin{equation}
\label{3.19}
\mathcal{Q}_I^2  \;  \equiv  \, ( \frac{\Delta T^I_2}{ T_0} )^2  \;  ,
\end{equation}
can be easily estimated from  Eq.~(\ref{3.4}):
\begin{equation}
\label{3.20}
\mathcal{Q}_I  \;  \simeq  \; ( \, 11.70 \; \pm \; 3.70 \, ) \; 10^{-6}  \;  .
\end{equation}
As a consequence   we get for the total quadrupole anisotropy:
\begin{equation}
\label{3.21}
\mathcal{Q}^2 = \mathcal{Q}_{A}^2 \; + \; \mathcal{Q}_{I}^2 \;  + \;  2 \, f(\theta_n,\phi_n,\phi_1,\phi_2) \,
\mathcal{Q}_{A} \mathcal{Q}_{I} \; ,
\end{equation}
where~\cite{Campanelli:2006,Campanelli:2007,Cea:2014}:
\begin{equation}
\label{3.22}
f(\theta_n,\phi_n,\phi_1,\phi_2)  \;  = \;  
             \frac{1}{4\sqrt{5}} \, [1 + 3 \cos(2 \theta_n) ]  \; + \;
                        \sqrt{ \frac{3}{10} } \, \sin (2 \theta_n) \, \cos (\phi_1 + \phi_n)  \; + \; 
               \sqrt{ \frac{3}{10} } \,  \sin^2 \theta_n \, 
               \cos ( \phi_2 + 2 \phi_n ) \; .
\end{equation}
Equations (\ref{3.21}) and  (\ref{3.22}) suggest that, if the space-time background metric is not isotropic, the quadruple anisotropy
may become smaller than the one expected in the standard isotropic $\Lambda$CDM cosmological model. 
To see this, we note that: 
\begin{equation}
\label{3.23}
 a_{20}  \; \simeq \;  + \,    \sqrt{\frac{\pi}{3}} \;   \mathcal{Q}_{I} \; + \; \frac{ 1}{6}  \, \epsilon^2   \sqrt{ \frac{\pi}{5} } \,
                  [1 + 3 \cos^2 (2 \theta_n) ]    \;  ,  
\end{equation}
\begin{equation}
\label{3.24}
 a_{21}  \;  = \;   + \, i \,  \sqrt{\frac{\pi}{3}} \; e^{i \phi_1} \;  \mathcal{Q}_{ I}  \; + \;
                      \, i \, \sqrt{ \frac{\pi}{30}}  \, \epsilon^2  \, e^{-i \phi_n}  \sin (2 \theta_n )  \;  , 
\end{equation}
\begin{equation}
\label{3.25}
a_{22}  \;    \simeq  \;  + \, \sqrt{\frac{\pi}{3}} \; e^{i \phi_2}  \; \mathcal{Q}_{I}  \;  + \; 
                                              \sqrt{  \frac{ \pi}{30} }  \, \epsilon^2   \; e^{-2 i \phi_n}  \sin^2 \theta_n   \; .
\end{equation}
Indeed, the quadrupole anomalies  can be accounted for if:
\begin{equation}
\label{3.26}
 a_{21}  \;  \approx \;   0  \;  \; \; , \; \;  |a_{20}|^2 \; \ll   2 \, | a_{22} |^2   \;  .
\end{equation}
It turned out that  Eq.~(\ref{3.26})  allowed us to determine the eccentricity at decoupling and constraint the polar angles 
of the symmetry axis~\cite{Cea:2014}:
\begin{equation}
\label{3.27}
 \epsilon^2  \; \simeq \;  \frac{ \sqrt{10} \, \mathcal{Q}_{ I}}{| \sin (2 \theta_n )|}   \;  , 
\end{equation}
and
\begin{equation}
\label{3.28}
 \theta_n  \; \simeq \;   \arctan \left ( \pm \frac{\sqrt{6}}{2} \; + \; 2 \right ) \; .
\end{equation}
Using  Eqs.~(\ref{3.21}), (\ref{3.27})  and Eq.~(\ref{3.28}) we reach an estimate of the eccentricity at decoupling a little bit smaller
than in Ref.~\cite{Cea:2014}: 
\begin{equation}
\label{3.29}
 e_{\rm dec} \;  \simeq  \;   ( 8.32 \, \pm \, 1.32) \, 10^{-3}   \; .
\end{equation}
As concern the quadrupole temperature anisotropy,  we have:
\begin{equation}
\label{3.30}
 \mathcal{Q}^2 \; \simeq \;   \frac{6}{5 \pi}   \;  | a_{22} |^2 \; \simeq \;  \frac{2}{5 } \, \mathcal{Q}^2_{I}  
\left [ 1 \, + \, \frac{  \sin^4 \theta_n}{  \sin^2 (2 \theta_n )} \, +  \, \frac{ 2  \sin^2 \theta_n}{  |\sin (2 \theta_n )|} \,   \cos (\phi_2 + 2 \phi_n) \right ] \; .
\end{equation}
The observed value of the quadrupole temperature anisotropy   is recovered if: 
\begin{equation}
\label{3.31}
\cos (\phi_2 + 2 \phi_n) \;  \simeq  \;   - \;  0.944  \; ^{+0.109}_{-0.056}    \; .
\end{equation}
To summarise,  we have shown that the anomalous low quadrupole temperature anisotropy can be reconciled  with observations
in the ellipsoidal universe by allowing a rather small  eccentricity at decoupling. On the other hand, in our previous paper~\cite{Cea:2014},
we found that the anisotropy of the metric contributes mainly at large scales affecting only the low-lying multipoles, at least for the temperature-temperature anisotropy correlations.  So that the ellipsoidal cosmological model can reproduce the observed quadrupole temperature
correlation without spoiling the excellent agreement of the standard cosmological model with the TT correlations for $\ell > 2$.
Finally, note that for the galactic coordinates $b_n,l_n$ of  the symmetry axis we have $b_n   \simeq     \pm  \,  17^{\circ} $ while the longitude $l_n$ is poorly constrained. It is worthwhile to observe that  the full mission Planck temperature data  exhibit  an 
anomalous mirror antisymmetry in  the direction $(b,l) \simeq (-17^{\circ}, 264^{\circ})$~\cite{Ade:2016b}.  Remarkably,  
the galactic coordinates of   the mirror antisymmetry direction are consistent with the allowed range of   $b_n,l_n$.  
This lead us to identify that direction with the direction of  the axis of symmetry. Therefore, we may safely assume:
\begin{equation}
\label{3.32}
 \theta_n  \; \simeq \;   \arctan \left (  2  \; - \; \frac{\sqrt{6}}{2}   \right ) \; \simeq  \; 73^{\circ}  \; \; \; , \; \; \;
 \phi_n  \; \simeq \;  264^{\circ}  \; .
\end{equation}
Now we turn on the large scale polarisation in the primordial cosmic background.  In our previous work~\cite{Cea:2010} we argued that
the ellipsoidal geometry of the universe induces sizeable polarisation signal at large scale without   needing  the CMB reionization mechanism.
We will assume, therefore, that early CMB reionization is negligible. As a consequence, as it is well 
known~\cite{Dodelson:2003,Mukhanov:2005},  at large scales  the primordial inflation induced cosmological
perturbations do not produce sizeable polarisation signal. In this case the polarisation of the temperature fluctuations
are fully accounted for by the anisotropic expansion of the universe. According to our previous discussion,  we have:
\begin{equation}
\label{3.33}
\Theta^{E}(\vec{k},\eta_0,\mu, \hat{p})  \; \simeq \;   \theta_p \, (1 \, - \, \cos^2 \theta_{\vec{p}  \vec{n}} )  \;  e^{- \, i \, k \,   \cos \theta_{\vec{k} \vec{ p}} \,  \eta_0}  \; ,
\end{equation}
where the superscript E  indicates that the temperature polarisation contributes only to the so-called E-modes. 
At large scales the main contributions to the polarisation temperature contrast functions  are for $k \, \simeq \, 0$:
\begin{equation}
\label{3.34}
\frac{\Delta T^{E}(\theta,\phi)}{ T_0} \;  \simeq \; \Theta^{E}(\vec{k} \simeq 0,\eta_0,\mu, \hat{p}) \; = \; \theta_p \, 
(1 \, - \, \cos^2 \theta_{\vec{p}  \vec{n}} ) \; = \;
 \frac{2}{3} \, \theta_p  \; - \; \frac{2}{3} \;  \theta_p \; P_2( \cos \theta_{\vec{p}  \vec{n}})   \; .  
\end{equation}
 It is evident from  Eq.~(\ref{3.34}) that  
the non-zero multipole coefficients  $a_{\ell m}^{ E}$ are for the monopole $\ell = 0$ and the quadrupole $\ell = 2$.
The monopole term indicates a non-zero average large scale  polarisation of the cosmic microwave background
that can be evaluated as:
\begin{equation}
\label{3.35}
(\frac{ \Delta T^{EE}_{0}}{T_0} )^2  \; \equiv \; \left ( \frac{1}{4 \pi} \; \int \; d\Omega \; \;  \frac{\Delta T^{E}(\theta,\phi)}{ T_0}  \right )^2
\;  \simeq \;  
 \frac{4}{9} \, \theta_p^2   \; .
\end{equation}
On the other hand, from Eq.~(\ref{3.34}) we easily obtain:
\begin{equation}
\label{3.36}
a_{2 m}^{ E}  \; \simeq \;   - \,  \frac{ 8 \pi}{15}  \, \theta_p   \;  Y^*_{2 m}(\theta_n,\phi_n)  \;  \; , 
\end{equation}
that implies:
\begin{eqnarray}
\label{3.37}
&& a_{20}^{E} \; \simeq \;  + \, \frac{ 1}{3}  \,  \theta_p  \,    \sqrt{ \frac{\pi}{5} } \;
                  [1 + 3 \cos^2 (2 \theta_n) ] \;  ,  \nonumber \\
&& a_{21}^{ E} \;  = \;  (a_{2,-1}^{A})^{*} \; \simeq \; + \, 2 \; i \;
                  \sqrt{  \frac{\pi}{30} }  \; \theta_p  \, 
                            e^{-i \phi_n}  \sin (2 \theta_n )  \;  , \nonumber \\
&& a_{22}^{ E} \; = \; (a_{2,-2}^{ A})^{*} \;  \simeq \; + \, 2 \,
                   \sqrt{  \frac{ \pi}{30} } \;  \theta_p  \,   
                  \; e^{-2 i \phi_n}  \sin^2 \theta_n  \; .
\end{eqnarray}
Using  Eq.~(\ref{3.37}) we may easily estimate the quadrupole TE and EE correlations. For the quadrupole TE correlation we get: 
\begin{equation}
\label{3.38}
( \frac{\Delta T^{TE}_{2}}{ T_0 } )^2 \; =  \;  \frac{3}{5 \,  \pi} \;
 \sum_{m=-2}^{m=+2} \,   a^{A}_{2 m} \,  (a^{E}_{2 m})^*  \;  =  \;  \frac{3}{5 \,  \pi} \; \left \{ a^{A}_{2 0} \,  a^{E}_{2 0}  \, + \,
2 \,  \mathcal{R}e \,  [ a^{A}_{2 1} \,  (a^{A}_{2 1})^*]  \, + \,
2 \,  \mathcal{R}e \,  [ a^{A}_{2 2} \,  (a^{E}_{2 2})^*]  \right \} \; ,
\end{equation}
where the $ a^{A}_{2 m}$'s are given by Eq.~(\ref{3.13}).  Proceeding as in Ref.~\cite{Cea:2014} one reaches the estimate:
\begin{equation}
\label{3.39}
( \frac{\Delta T^{TE}_{2}}{ T_0 } )^2  \;  \simeq  \; \frac{4 }{5 \,  \sqrt{10}} \; \theta_p \; \mathcal{Q}_I  \;  \sin^2 \theta_n \;
 \left [ \cos (\phi_2 + 2 \, \phi_n) \; + \; \frac{  \sin^2 \theta_n }{ | \sin (2 \, \theta_n) |}   \right ] \;.
\end{equation}
For the quadrupole EE correlation  we get:
\begin{equation}
\label{3.40}
( \frac{\Delta T^{EE}_{2}}{ T_0 } )^2 \; =  \;  \frac{3}{5 \,  \pi} \;
 \sum_{m=-2}^{m=+2}  | a^{E}_{2 m} |^2 \; \simeq \; \frac{16}{75} \;   \theta_p^2  \; .  
\end{equation}
From Eqs.~(\ref{3.35}),  (\ref{3.39})  and  (\ref{3.40})  it is   straightforward to obtain:
\begin{equation}
\label{3.41}
(\Delta T^{TE}_{2})^2   \;  \simeq  \;  8.28  \; \pm  \; 3.70   \;   \mu \, K^2  \; ,
\end{equation}
\begin{equation}
\label{3.42}
( \Delta T^{EE}_{0} )^2  \;  \simeq  \;  1.26  \; \pm  \;  0.80  \;  \mu \, K^2  \; ,
\end{equation}
\begin{equation}
\label{3.43}
(\Delta T^{EE}_{2})^2   \;  \simeq \;  0.60 \; \pm \; 0.38  \;   \mu \, K^2  \; .
\end{equation}
\subsection[]{Comparison with the Planck 2018 data}
\label{S3.1}
%
\begin{figure}
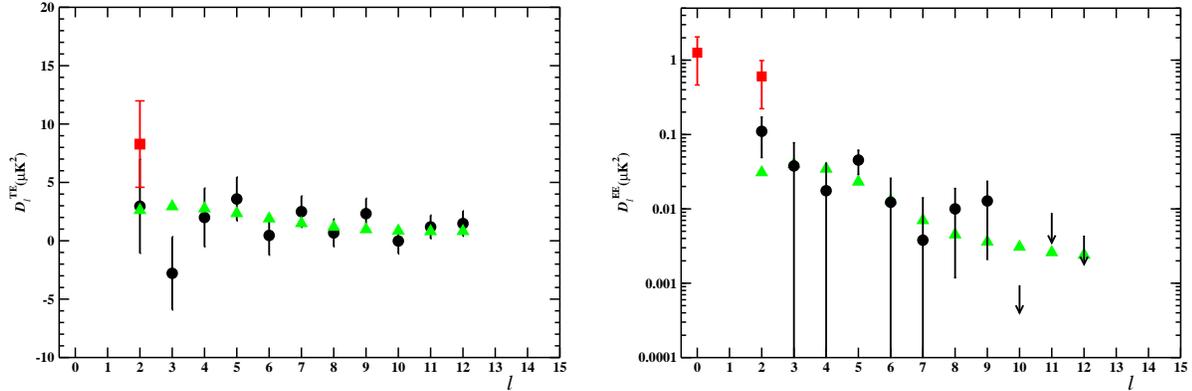

\begin{center}
\hspace{-0.6cm}
\includegraphics[width=0.465\textwidth,clip]{fig1a.eps}
\hspace{0.3cm}
\includegraphics[width=0.48\textwidth,clip]{fig1b.eps}
\caption{
(color online) The TE (left panel)  and EE (right  panel) temperature correlations at large scales $\ell \le 12$.  Full (black) points are the latest Planck data.
The arrows indicate the  upper limits at   68 \% confidence level. The full (green)  triangles are the  TE and TT temperature
correlations of the $\Lambda$CDM best fit. The  full (red) squares are our results Eqs.~(\ref{3.41}), (\ref{3.42}) and  (\ref{3.43}). }
\label{Fig1}
\end{center}
\end{figure}
 Figure~\ref{Fig1}  summarises what we have achieved in this work. Indeed, in Fig.~\ref{Fig1} we report 
TE  (left panel)  and EE (right  panel) temperature correlations at large scales  $\ell \le 12$.  It is worthwhile to observe
that measuring the CMB polarisation signals at large angular scales is challenging due to the contamination from foreground sources.
In general the main sources of foreground polarisation signals are synchrotron radiation, dust emission and thermal bremsstrahlung.
To extract the underlying CMB signal, the foreground signal must be removed. The removal processes form a significant part of the
CMB data analysis becoming increasingly important for the large-scale polarisation signals. The Planck team used component separation
techniques to produce various maps of foregrounds. Foreground removal techniques make specific assumptions about the properties
of foregrounds in temperature and in polarisation. To obtain a characterisation of the foregrounds that is independent of  the
component-separation methods, the Planck Collaboration, after subtracting from the observed power spectra the CMB contribution 
determined by using the Planck 2015 $\Lambda$CDM  model~\cite{Ade:2016a}, performed a power-law fits to the power spectra
over the multipole range $40 \; \leq \; \ell \; \leq \; 600 $.  After that,  the extrapolation of the fitted power laws to the low multipoles 
was compared to the data points at $\ell \; < \; 40$ not used in the fit. It turned out that the extrapolation of the power laws to low
multipoles was not alway close to the data points~\cite{Akrami:2018e}. This quantifies the challenge of the component-separation 
procedure that is required for measuring the low-$\ell$ primordial CMB E-modes. We see, then, that on the largest angular scales
the Planck's results for the primordial CMB polarisation remain contaminated by astrophysical foreground and unknown systematic
errors.  The latest Planck data for the large-scale TE and EE power spectra are displayed in Fig.~\ref{Fig1} as full points~\cite{Planck:2018}.
The data are  affected by a rather large statistical uncertainties mainly due to the problematic component-separation procedures.
Nevertheless, one can safely affirm that there is an evidence of primordial CMB polarisation signal at large scales. It is widely believed
that the primordial CMB polarisation signal at large scales is due to the reionization processes.  Even though the process by which
the universe become fully reionized is still not well characterised,  the reionization process is estimated to be complete at the redshift
of reionization $z_{re}$. The best fit $\Lambda$CDM model  gave $ z_{re} \; = \; 7.67 \; \pm 0.67$ and Thomson
optical depth $  \tau  \; = \; 0.0544 \; \pm 0.0073$ (see Table 2 in \cite{Aghanim:2018b}). The resulting TE and EE multipole correlations
are reported in  Fig.~\ref{Fig1} (full triangles). Looking at  Fig.~\ref{Fig1} it seems that the best fit model is in reasonable agreement
with the Planck data within the rather large statistical errors. However, due to the measurement uncertainties there is no a clear evidence
of the E-mode reionization bumps that should be around $\ell \; \simeq \; 3 \; - \; 4$ for a Thompson scattering optical depth 
$  \tau  \; \simeq \; 0.055$. On the other hand, we said  that  the ellipsoidal geometry of the
universe induces sizeable polarisation signal at large scales even without invoking the reionization scenario.
Our main results, given by  Eqs.~(\ref{3.41}), (\ref{3.42}) and  (\ref{3.43}), are compared to the Planck data in 
Fig.~\ref{Fig1} (full squares). We see that our estimate of the quadrupole TE correlation is consistent with observations
within the statistical errors. As concern the quadrupole EE correlation, our result  Eq.~(\ref{3.43}) agrees with the data point at
the same confidence level as the best fit model.  Moreover, in Ref.~\cite{Cea:2014} we already argued that in our cosmological model
 the polarisation  of the cosmic microwave background (without reionization) at the present time is essentially 
that produced around the time of recombination and, due to the finite thickness of the last scattering surface, 
it should be confined up to multipoles   $\ell \lesssim 10$.  In fact, this is confirmed in Ref.~\cite{Pontzen:2007}  where
the authors   have  derived the  radiative transfer equation  in  homogeneous  anisotropic  universes in the  Bianchi VII$_h$ case arguing
that qualitatively the level of polarisation induced by the spatial anisotropy  of the metric  should hold also for  Bianchi I
anisotropic universes. Therefore, in our cosmological anisotropic model  the  polarisation signal  should smoothly  extend
up multipoles $\ell \simeq 10$ and after that fall off very rapidly, in qualitative agreement with the Planck data.
From the TE and EE power spectrum at large angular scales we see that the Bianchi-induced polarisation can mimic
the effects of the early reionization of the standard cosmological model. Unfortunately, the Planck 2018 data on the large-scale 
polarisation are not yet able to distinguish  between these two model.  Nevertheless, a clear signature of the
 ellipsoidal universe model resides on the fact that, at variance of the standard reionization scenario, there is a 
 non-zero average temperature polarisation. Our estimate of the average polarisation (monopole EE correlation) given by Eq.~(\ref{3.42})  
 is displayed in  Fig.~\ref{Fig1}, right  panel. In principle, the monopole EE correlation can be easily measured by averaging
 the polarisation signal over all directions. However, the lack of adeguate theoretical models able to explain the
 foreground polarisation at large scales prevents, at moment,  a reliable estimate of the average polarisation. 
In fact, the eventual presence of an average temperature polarisation could be misinterpreted 
as foreground emission leading  to an underestimate of the cosmic microwave background polarisation signal. 
\section{Conclusions}
\label{S4}
The final results on the CMB anisotropies by the Planck Collaboration     are confirming the cosmological  Lambda Cold Dark Matter  model 
to the  highest level of accuracy.  Nevertheless, at large angular scales there are still anomalous features in CMB anisotropies.
Actually, the most evident discrepancy  resides in the quadrupole TT correlation.  The latest observed quadrupole TT correlation is:
\begin{equation}
\label{4.1}
(\Delta T^{TT}_{2})^2   \;  = \;  225.90   \; ^{+ 533.06}  _{-132.37}   \;  \;  \mu \, K^2  \; ,
\end{equation}
where the estimated errors take care of the cosmic variance. On the other hand,  the 'TT,TE, EE + low E + lensing'  best fit
 $\Lambda$CDM  model to the Planck 2018 data gave:
\begin{equation}
\label{4.2}
( \Delta T^{TT}_{2} )^2_{ \Lambda CDM} \;  =  \;  1016.73  \;   \mu \, K^2  \; ,
\end{equation}
that differs from the observed value by about two standard deviations.
It has been suggested~\cite{Campanelli:2006,Campanelli:2007} that,  if one assumes that the large-scale spatial geometry of our universe 
is slightly anisotropic, then the quadrupole amplitude can be drastically reduced without affecting higher multipoles of the angular power spectrum of the temperature anisotropies~\cite{Cea:2010,Cea:2014}. In the present paper, that relies  heavily on our previous work, 
we performed a stringent comparison with the latest CMB data from the Planck Collaboration. As in previous papers we established that the low quadrupole temperature correlation, detected by WMAP and confirmed by the Planck satellite, could be accounted for  if the geometry of the universe is plane-symmetric with  eccentricity at decoupling of order $10^{-2}$.  In fact, we  fixed   the eccentricity at decoupling, Eq.~(\ref{3.29}),
and the polar angles of the direction of the axis of symmetry, Eq.~(\ref{3.32}),  such that the quadrupole TT correlation matches exactly the
observed value Eq.~(\ref{4.1}).  As a consequence the ellipsoidal universe model seems to compare  a little bit better than the standard
 cosmological  Lambda Cold Dark Matter  model. On the other hand, it is known since long time that anisotropic cosmological model
 could induce sizeable large-scale CMB polarisation. Indeed, we already argued that in the ellipsoidal universe model there is
 a sizeable polarisation signal at scales $\ell \; \lesssim \; 10$.  Moreover,  we showed that the quadrupole
 TE and EE correlations in the ellipsoidal universe are in reasonably agreement  with the Planck 2018 data. 
 Therefore, we are led to conclude that the proposal of the ellipsoidal universe cosmological model is still a viable alternative 
  to the standard cosmological model.  Finally, we suggested that a reliable estimate
 of the average large-scale polarisation  by the Planck Collaboration could  confirm or reject the ellipsoidal  universe
 proposal.


\begin{thebibliography}{}
%
%
\bibitem{Aghanim:2018a}
 N. Aghanim   et al.,  Planck Collaboration, 
 Planck 2018 results. III. High Frequency Instrument data processing and frequency maps, 
  arXiv:1807.06207 [astro-ph.CO] (2018)
%
%
\bibitem{Aghanim:2018b}
N.  Aghanim   et al.,  Planck Collaboration,  
Planck 2018 results. VI. Cosmological parameters,
arXiv:1807.06209 [astro-ph.CO] (2018)
%
%
\bibitem{Aghanim:2018c}
N.  Aghanim  et al.,  Planck Collaboration,
Planck 2018 results. VIII. Gravitational lensing,
  arXiv:1807.06210 [astro-ph.CO] (2018)
%
%
\bibitem{Aghanim:2018d}
N.  Aghanim  et al.,  Planck Collaboration, 
Planck 2018 results. XII. Galactic astrophysics using polarized dust emission,
arXiv:1807.06212 [astro-ph.GA] (2018)
%
%
\bibitem{Akrami:2018a}
 Y. Akrami  et al.,  Planck Collaboration, 
 Planck 2018 results. I. Overview and the cosmological legacy of Planck,
 arXiv:1807.06205 [astro-ph.CO] (2018)
%
%
\bibitem{Akrami:2018b}
 Y. Akrami  et al.,  Planck Collaboration, 
 Planck 2018 results. II. Low Frequency Instrument data processing,
 arXiv:1807.06206 [astro-ph.CO] (2018)
%
%
\bibitem{Akrami:2018c}
 Y. Akrami   et al.,  Planck Collaboration, 
 Planck 2018 results. IV. Diffuse component separation, 
  arXiv:1807.06208 [astro-ph.CO] (2018)
%
%
\bibitem{Akrami:2018d}
 Y. Akrami  et al.,  Planck Collaboration, 
 Planck 2018 results. X. Constraints on inflation, 
   arXiv:1807.06211 [astro-ph.CO] (2018)
%
%
\bibitem{Akrami:2018e}
 Y. Akrami  et al.,  Planck Collaboration, 
 Planck 2018 results. XI. Polarized dust foregrounds,
 arXiv:1801.04945 [astro-ph.GA] (2018)
%
\bibitem{Campanelli:2006} 
L. Campanelli, P.  Cea  and  L. Tedesco, 
Ellipsoidal Universe Can Solve the Cosmic Microwave Background Quadrupole Problem,
Phys. Rev. Lett.   {\bf 97}, 131302  (2006) 
%
\bibitem{Campanelli:2007} 
L. Campanelli, P.  Cea  and  L. Tedesco, 
Cosmic microwave background quadrupole and ellipsoidal universe,
Phys. Rev. D {\bf 76} , 063007 (2007)
%
\bibitem{Peebles:1993} 
P.~J.~E.  Peebles,  Principles of Physical Cosmology, Princeton University Press, Princeton (1993)
%
 \bibitem{Bianchi:1898} 
 L. Bianchi,  Sugli spazi a tre dimensioni che ammettono un gruppo continuo di movimenti,
 Mem. Mat.  Fis.  Soc. It. Sci., serie {\bf III},  {\bf Tomo  XI}, 267 (1898);  
 English translation:   On three dimensional spaces which admit a group of motions,
Gen. Rel. Grav. {\bf 33},  2157  (2001)
%
\bibitem{Rees:1968}
M.~J. Rees,  
Polarization and Spectrum of the Primeval Radiation in an Anisotropic Universe,
Astrophys. J.  {\bf 153} , L1 (1968)
%
\bibitem{Negroponte:1980} 
J. Negroponte  and  J.  Silk,
Polarization of the Primeval Radiation in an Anisotropic Universe,
 Phys. Rev. Lett. {\bf 44}, 1433 (1980)   
%
\bibitem{Basko:1980}
M.~M. Basko  and  A.~G. Polnarev,  
Polarization and anisotropy of the relic radiation in an anisotropic universe,
Monthly Not. R. Astron. Soc. {\bf 191},  207 (1980)
 %
\bibitem{Pontzen:2007}  
A. Pontzen  and A.  Challinor,  
Bianchi model CMB polarization and its implications for CMB anomalies,
Monthly Not. R. Astron. Soc. {\bf  380}, 1387 (2007)
%
\bibitem{Ade:2015}
P.A.R. Ade  et al., BICEP2/Keck and Planck Collaborations,
Joint Analysis of BICEP2/Keck Array and Planck Data,
 Phys. Rev. Lett. {\bf 114}, 101301 (2015)
%
\bibitem{Cea:2010}  
P.  Cea,  
On the large-scale cosmic microwave background polarization,
Monthly Not. R. Astron. Soc. {\bf  406}, 586 (2010)
%
\bibitem{Cea:2014}   
P. Cea, 
The ellipsoidal universe in the Planck satellite era,
Monthly Not. R. Astron. Soc.  {\bf  441}, 1646 (2014)
%
\bibitem{Bennett:2013} 
C.L. Bennett   et al., 
Nine-year Wilkinson Microwave Anisotropy Probe (WMAP) Observations: Final Maps and Results,
Astrophys. J. Suppl. {\bf 208}, 20 (2013)
%
\bibitem{Hinshaw:2013} 
 G.F. Hinshaw   et al., 
Nine-year Wilkinson Microwave Anisotropy Probe (WMAP) Observations: Cosmological Parameter Results,
 Astrophys. J. Suppl. {\bf  208}, 19 (2013)
%
%
\bibitem{Dodelson:2003}  
S. Dodelson, {\it  Modern Cosmology},  Academic Press, San Diego, California (2003)
%
\bibitem{Mukhanov:2005}  
 V. Mukhanov, {\it Physical Foundation of Cosmology},  Cambridge University Press, New York (2005)
%
\bibitem{Chandrasekhar:1960} 
S. Chandrasekhar,  {\it Radiative Transfer}, Dover Publications, New York (1960)
%
\bibitem{Planck:2018} 
Planck 2018: https://www.cosmos.esa.int/web/planck
%
\bibitem{Fixsen:2009}
D. J.  Fixsen,   
The Temperature of the Cosmic Microwave Background,
Astrophys. J.  {\bf  707}, 916 (2009)
%
\bibitem{Ade:2016b} 
P. A. R. Ade  et al., Planck Collaboration, 
Planck 2015 results. XVI. Isotropy and statistics of the CMB,
Astron. Astrophys.   {\bf 594}, A16 (2016)
 %
\bibitem{Ade:2016a} 
 P. A. R.  Ade. et al., Planck Collaboration, 
Planck 2015 results. XIII. Cosmological parameters,
Astron. Astrophys.   {\bf 594}, A13 (2016)
 %
%
%
\end{thebibliography}
\end{document}